# opp/ai: Optimistic Privacy-Preserving AI on Blockchain

Cathie So

ORA*, cathie@ora.io

KD Conway

ORA

Xiaohang Yu

ORA

Suning Yao

ORA

Kartin Wong

ORA

The convergence of Artificial Intelligence (AI) and blockchain technology is reshaping the digital world, offering decentralized, secure, and efficient AI services on blockchain platforms. Despite the promise, the high computational demands of AI on blockchain raise significant privacy and efficiency concerns. The Optimistic Privacy-Preserving AI (opp/ai) framework is introduced as a pioneering solution to these issues, striking a balance between privacy protection and computational efficiency. The framework integrates Zero-Knowledge Machine Learning (zkML) for privacy with Optimistic Machine Learning (opML) for efficiency, creating a hybrid model tailored for blockchain AI services. This study presents the opp/ai framework, delves into the privacy features of zkML, and assesses the framework's performance and adaptability across different scenarios.

**Additional Keywords and Phrases:** Blockchain, Machine Learning, Zero Knowledge Proofs, Fraud Proofs

## 1 INTRODUCTION

In the dynamic digital environment, the fusion of Artificial Intelligence (AI) and blockchain technology is significantly transforming our interaction with and utilization of information. AI, known for its sophisticated data processing and decision-making capabilities, combined with blockchain's decentralized and secure ledger system, is pioneering new avenues in the digital domain. This collaboration has led to the emergence of "Onchain AI," a concept that promises to deliver decentralized, secure, and efficient AI services within the blockchain networks [1,10].

However, implementing AI computations directly on the blockchain presents challenges, notably due to the high computational costs. For instance, executing a basic operation like matrix multiplication with 1000 × 1000 integers on the Ethereum blockchain would require over 3 billion gas [23], surpassing the platform's block gas limit [22]. As a result, many applications opt for off-chain computations on centralized servers, uploading only the results to the blockchain. This approach, while functional, compromises the decentralization principle, raising security concerns and undermining the trust and transparency blockchain aims to provide.

An innovative solution to this challenge is Zero-Knowledge Machine Learning (zkML), which leverages zero-knowledge proofs (ZKPs) to protect confidential data and model parameters during training and inference, thus addressing privacy issues and reducing blockchain's computational load [24]. zkML, however, faces its own set of challenges, particularly the high costs and computational demands of generating proofs [18], making it less feasible for large-scale AI applications.

---

* formerly known as HyperOracle

To address the limitations of zkML, the concept of Optimistic Machine Learning (opML) on the blockchain has been introduced [3]. Unlike zkML, opML utilizes a fraud-proof system to ensure the correctness of ML results, maintaining an optimistic assumption about the validity of submitted results. This system involves a challenge period where results can be disputed, and if necessary, a detailed arbitration process is conducted on the blockchain, requiring minimal computational resources. This approach, exemplified by rollup systems like Optimism and Arbitrum [8], offers a more efficient and scalable solution for integrating ML with blockchain technology, paving the way for secure, decentralized, and transparent AI services on the blockchain.

The optimistic approach to machine learning on the blockchain, while offering a solution to the high computational costs associated with zkML, does raise concerns regarding data privacy. In this approach, all data must be publicly available for challengers to validate the submitted results, which starkly contrasts the privacy-preserving capabilities of zkML. The public nature of data in the optimistic approach may hinder its adoption in situations where privacy is critical, such as in handling sensitive health or financial information. This highlights a fundamental trade-off in blockchain-based machine learning solutions between ensuring computational efficiency and maintaining privacy.

The Optimistic Privacy-Preserving AI (opp/ai) framework represents an innovative approach to addressing the challenges of privacy and computational efficiency in blockchain-based machine learning systems. This framework is designed to mitigate the high cryptographic overhead associated with zkML while making strategic trade-offs in privacy to ensure a balanced solution. The opp/ai framework is underpinned by prohibitive costs to guarantee security, addressing some of the inherent privacy limitations of zkML, particularly in the context of sensitive applications like finance, government, and health.

Key Features of the opp/ai framework:
- **Reduced Cryptographic Overhead**: By making small concessions in privacy, the opp/ai framework significantly reduces the computational and resource-intensive demands of zkML. This is achieved without compromising the core objective of preserving privacy during ML computations on the blockchain.
- **Security Guarantees**: The framework ensures security through game theory. Prohibitive costs act as a deterrent against attacks, ensuring that any attempt to breach privacy or integrity is economically unfeasible.
- **Learning from DNN Partitioning and Split Learning**: The opp/ai framework incorporates strategies from Deep Neural Network (DNN) partitioning [9,20] and split learning [12]. This involves dividing the ML model into segments that can be processed separately, reducing the amount of data that needs to be shared and processed optimistically. This partitioning approach helps maintain privacy by limiting the exposure of data and computations.

The contribution of this work is summarized as follows:
- We propose opp/ai: Optimistic Privacy-Preserving AI on blockchain. Compared to zkML or opML alone, opp/ai provides a framework to balance the trade-offs between privacy and computational efficiency strategically. This framework leverages the strengths of both zkML's privacy-preserving techniques and opML's efficiency in computation, creating a hybrid model that aims to optimize both aspects for blockchain-based AI applications.

- We analyze the privacy-preserving properties of zkML. While ZKPs enable the verification of computations without exposing the underlying data, it is important to note that using ZKPs does not inherently guarantee data privacy in the machine learning setting. We evaluate these properties in Section 3 Security Analysis.

The organization of this paper goes as follows. In Section 2, we introduce the architecture and the workflow of opp/ai. In Section 3, we analyze the privacy-preserving properties of zkML on both the input privacy and model privacy aspects. In Section 4, we present the theoretical performance of the opp/ai framework, along with actual benchmarking results. Section 5 discusses the versatility of opp/ai, its adaptability to different models, and its various use cases. The paper concludes with a final remark in Section 6.

## 2 OVERVIEW

### 2.1 Architecture

opp/ai applies architecture from opML and zkML for different parts of an ML inference. opML has three integral components [3]:

- **Fraud Proof Virtual Machine (FPVM).** FPVM can trace any instruction step of a stateless program and prove it on Layer 1 blockchain (L1).
- **Machine Learning Engine.** The machine learning engine is incredibly efficient and has been designed to cater to both native execution and fraud-proof scenarios. This engine ensures that machine learning tasks are executed quickly and accurately, while guaranteeing the results' consistency and determinism.
- **Interactive Dispute Game.** The dispute game will resolve a disputed instruction using on-chain FPVM.

zkML has two integral components:

- **Prover.** The prover generates ZKPs, such as zk-SNARKs [17], to demonstrate the validity of the ML inference.
- **On-chain Verifier.** The verifier smart contract efficiently validates the zero-knowledge proofs provided by the prover [5].

### 2.2 Workflow

We adopt various opp/ai workflows based on the privacy-preserving scenario we intend to achieve. The parties involved in the workflow are defined as follows:

- **Prover.** The prover is responsible for generating the proof of computation for the submodel(s) running in zkML.
- **Requester.** The requester initiates an opML task and sends it to the submitter (and challengers).
- **Submitter.** The submitter commits the results of the submodel(s) running in opML onto the blockchain.
- **Challengers.** The challengers check these results. In case of any dispute, a challenger and the submitter enter the dispute game together.

Assuming we can partition an ML model $f$ into $2n$ submodels, for $1 \leq i \leq n$, $f_i^o$ denotes the $i$-th submodel running in opML, and $f_i^z$ denotes the $i$-th submodel running in zkML. We have

$$f = f_n^o \circ f_n^z \circ \ldots \circ f_2^o \circ f_2^z \circ f_1^o \circ f_1^z$$

It is important to note that any of these submodels could be "empty," such that $f_i^{\{o,z\}}(x) = x$. Hence, the above expression is in its most general form.

To achieve model privacy, the proofs of all $f_i^z$'s should have its model input set as public input, while the model parameters or weights should be kept private or hashed. The workflow for model privacy unfolds as follows:

1. All $f_i^o$'s are published as a public model available to all potential submitters/challengers.
2. The prover, in this case, the model provider, receives the model input from the end user and computes the proofs of all $f_i^z$'s, submitting them to the on-chain verifiers while also requesting to initiate an opML service task on $f_i^o$'s.
3. The submitter performs the opML service task, taking the result from each $f_i^z$ as input to each $f_i^o$, and commits the results on the blockchain.
4. The challengers validate the results and will start a dispute process with the submitter if they find any of them inaccurate. The dispute process will only apply to the specific $f_i^o$ being challenged.
5. The smart contract facilitates arbitration to resolve any dispute, providing a conclusive and final resolution. After a defined "challenge period," the results will be confirmed [3].

Similar to the case with input privacy, the prover and the submitter could be the same entity to expedite the submission of the initial results.

## 3 SECURITY ANALYSIS

In this section, we will analyze the security of zkML technology in the context of the opp/ai framework. Our analysis indicates that the vulnerabilities found in zkML are not limited to opp/ai but are inherent to zkML technology as a whole. We have investigated potential security weaknesses and attack vectors, and our simulations have provided empirical evidence of how these vulnerabilities can be practically exploited. The main takeaway from this analysis is that the security issues arise from the zkML technology itself.

In the context of preserving model privacy, the single prover trust assumption posits that only one trusted entity, the prover, has access to the model weights. All end users interact with this prover to generate zkML proofs without direct access to the model weights themselves. This approach is designed to protect the confidentiality of the model while still enabling users to benefit from its predictive capabilities. However, this setup introduces a potential vulnerability: an attacker could run a large number of inferences by paying for the service and use the results of these inferences to train a proxy or heuristic model [21]. This proxy model could approximate the behavior of the original model, effectively circumventing the privacy measures in place. The security of the system in such a scenario relies on the cost of inferences rather than the strength of the ZKP itself. Therefore, the opp/ai framework can achieve a similar level of model privacy by concealing submodel weights and trading partial privacy for better computational performance.

We attempt to model the cost of an attack based on the assumptions and variables as follows:

- It is assumed that the number of inferences required to retrieve model weights when the architecture is known increases linearly proportionally with the size of the model. This idea has been well-established in the field of regression analysis [14] and we are using it as a rule of thumb for neural networks.
- $p$ = the proportion of the model running in zkML under opp/ai
- $x$ = model size, in units of model parameters, constraints (in R1CS system such as *circom* [27]), rows (in *halo2* [28]), or opcodes/instructions (in *zkGo* [29] or *zkWasm* [30])
- $c$ = the cost per inference set by the prover
- $n$ = the number of inferences required to reconstruct the model per unit of $x$

We have

$$\text{attack cost} = c \times n(1-p)x$$

based on the above assumptions and definitions.

Therefore, there are two ways to safeguard or prevent the model from such an attack:

1. To limit the maximum number of inferences on the model, such that the attacker can never make as many as $n(1-p)x$ inferences. This approach is suitable for use cases such as AI-generated content (AIGC) non-fungible tokens (NFTs) [15], where the scarcity of inference outputs is desired.
2. Adjust $c$ to be inversely proportional to $(1-p)$, such that the attack cost is constant regardless of how much of the model is run in zkML.

Practically, the cost of inference should be determined by the underlying cost of proof generation and opML computation, outrunning the cost of the attacker collecting their own data and training their own model.

In addition to the general model privacy considerations above, it is important to highlight the framework's applicability to fine-tuned models. Specifically, opp/ai can be utilized to conceal the fine-tuning weights of models where the majority of the weights are already publicly available. This is particularly relevant for open-source models that have been fine-tuned for specialized tasks. For instance, the LoRA weights in the attention layers of the Stable Diffusion model [7,16] can be protected using our framework, as demonstrated in our proof of concept in Section 4.3. This capability is crucial for preserving the proprietary enhancements made to publicly shared models, ensuring that while the base model remains accessible, the unique adaptations that provide competitive advantage remain confidential.

On the other hand, to ensure user privacy, zkML proof generation should ideally occur on local devices, where model weights are distributed. However, this can leave systems vulnerable to reverse-engineering attacks, where adversaries might reconstruct inputs from model outputs, compromising privacy*. This risk is especially high in models that maintain high input data dimensionality. Given these challenges with current zkML technology, the opp/ai framework will concentrate on model parameter privacy rather than input privacy.

This focus on model parameter privacy complements the framework's ability to protect fine-tuning weights of widely available models, thus securing competitive advantages while the base model remains open-source.

## 4 BENCHMARK

We suggest taking two approaches to assess the efficacy of the opp/ai framework for computation. First, develop a theoretical model to gauge the computational cost. Second, conduct a practical evaluation of the framework's performance on a real machine learning model. This will help determine the potential efficiency gains achievable by implementing the opp/ai framework and compare it to pure zkML.

### 4.1 Theoretical performance

We assume that the computational cost, in terms of time or memory required, increases linearly with the model size $x$. However, performing the same computation in zkML is currently estimated to be 1000 times more expensive [18].

Hence, we can assume that the computation cost is dominated by the zkML overhead and correlated to the proportion of the model running in zkML under opp/ai. In simple words, if we decide to run a part of the model in opML rather than zkML, we can save computation costs directly correlated to that part.

---

\* See Figure A.1 in the appendices for an illustration of potential reconstruction attacks on input data.

## 4.2 Actual performance

We benchmark two zkML frameworks, *keras2circom* [19] and *EZKL* [31], to compare proof generation time and prover maximum memory usage. A simple MNIST convolutional neural network [32] (Figure 1) is converted to circuits in each framework, and the *wall clock time* and maximum prover *memory consumption* are measured while running the prover. Similar to previous work [18], circuit compilation and setup, verifier runtime, and generated proof size are omitted. The benchmark primarily measures the computation costs associated with proof generation.

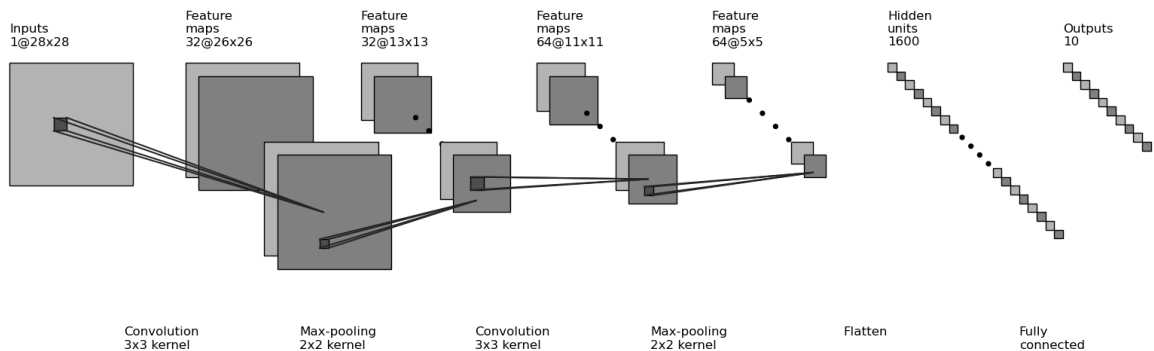

Figure 1: Architecture of the simple MNIST convolutional neural network*

### 4.2.1 Methodology

To benchmark the model, we have removed the last layer of softmax activation, as it can cause calibration issues and introduce unnecessary overhead. While the opp/ai framework allows for flexibility to apply zkML or opML to any part of the model, we assume that the first layers of the model are run in zkML and the rest of the model is run in opML. We have set the model input to be public and the model parameters to be private in both frameworks.

We extract a model from the first layer up to the *i*-th layer and export it to *h5* format for *keras2circom* and ONNX format for *EZKL*. The *flatten* and *dropout* layers are not iterated since they do not add constraints (in *circom*) or rows (in *halo2*). To test the model, we take one of the images from the original MNIST dataset [11] as an input.

For *circom*, we compile the circuit, generate the witness, and set up the verification key under the Groth16 scheme [6] in *snarkjs* [33]. The model size is defined in units of constraints in R1CS (Rank-1 Constraint Systems). For *EZKL*, we generate and calibrate the settings based on the sample input, compile the circuit, set up the verification key, and generate the witness. The model size is defined in units of rows in the halo2 circuits. We then measure the proof generation time and memory usage for each framework. We use *TensorFlow* [4] to run the non-zkML part of the inference as a heuristic to estimate the computation time of opML. This time is then added to the total computation time. As the memory usage of opML is generally much lower than that of zkML, we have omitted its measurement.

To compare two frameworks on the same scale and validate our theoretical prediction, we have transformed the measurements into percentages of total time, maximum memory usage, and model size, respectively. All benchmarks were run on AMD EPYC 7R13 processors (AWS's r6a.16xlarge instances) with 512GB of RAM.

---

* This figure is generated by adapting the code from https://github.com/gwding/draw_convnet.

*4.2.2 Results**

Figure 2 presents the time taken to generate proofs in the opp/ai framework, comparing the framework with the *circom*-only and *EZKL*-only setups. For the opp/ai with *circom* setup, the first bar on the left shows that at a 46% proportion of the model running in *circom* ($p = 0.46$), the proof generation time is about 5 minutes (approximately 42% of the maximum time observed). When the entire model runs in zkML, the time increases to around 11 minutes. In the opp/ai with *EZKL* setup, the proof generation is 36 seconds at a 16% zkML proportion (approximately 19% of the maximum time; $p = 0.16$) and a little over 3 minutes at full zkML proportion.

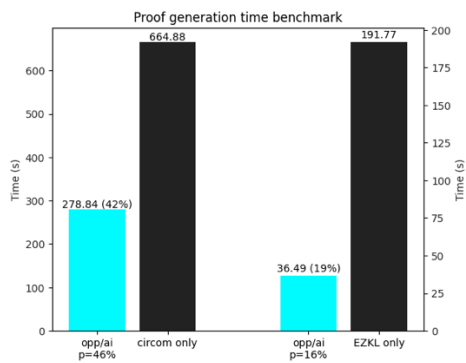

Figure 2: Proof generation time comparison between using zkML only and using opp/ai

Figure 3 compares the memory consumption in the same setups. In the opp/ai with *circom* setup, memory usage at a 46% zkML proportion is approximately 85GB (62% of the maximum observed memory usage), escalating to about 138GB when the full model is in zkML. For the opp/ai with *EZKL* setup, memory usage is around 3GB at a 16% zkML proportion (13% of the maximum) and 23GB when running fully in zkML. The raw data of Figures 2 and 3 can be found in Table 1.

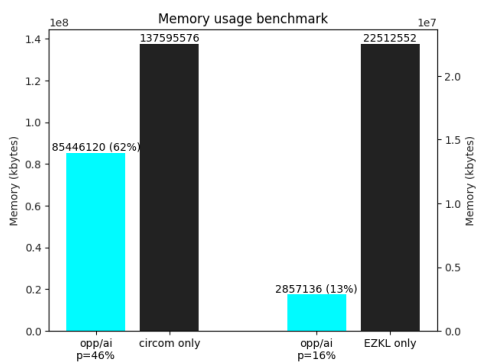

Figure 3: Maximum memory usage comparison between using zkML only and using opp/ai

The results above demonstrate that running a smaller portion of the model in zkML can significantly reduce both memory usage and proof generation time, resulting in substantial efficiency gains. The computation cost is almost directly proportional to $p$, the proportion of the model running in zkML under opp/ai. In other words, if we decide to run a part of the model in opML rather than zkML, the computation costs we save are directly proportional to the amount of the model run in opML instead of zkML. This improved efficiency is essential for optimizing the framework for various use cases, where the balance between zkML and opML components can be adjusted to achieve the desired privacy and computational efficiency.

---

* The complete benchmarking scripts and results can be accessed at https://github.com/socathie/oppai-benchmark/tree/main/mnist.

Table 1: Raw benchmarking data

| Model Layer | circom | | | | EZKL | | TensorFlow |
|---|---|---|---|---|---|---|---|
| | Constraints | Time (mm:ss) | Memory (kB) | Rows | Time (mm:ss) | Memory (kB) | Time (ns) |
| Conv2D | 11378432 | 04:38.80 | 85446120 | 184408 | 00:33.20 | 2869864 | 40334380 |
| MaxPooling2D | 16867552 | 08:44.18 | 129358852 | 224968 | 00:36.45 | 2857136 | 39514066 |
| Conv2D | 23101472 | 10:16.60 | 136911572 | 1373720 | 03:08.11 | 22516096 | 39434894 |
| MaxPooling2D | 24725472 | 10:50.14 | 138164812 | 1385048 | 03:08.13 | 22510672 | 38969498 |
| Dense | 24741472 | 11:04.88 | 137595576 | 1392273 | 03:11.77 | 22512552 | 38995473 |

### 4.3 Extrapolation to Stable Diffusion model*

One key contribution of opp/ai is the possibility to hide fine-tuned model weights from an open-sourced model, such as the LoRA weights in the Stable Diffusion model [7,16]. Under opp/ai, it is possible to apply zkML only on the attention layers of the U-Net model that contain the LoRA weights, while running opML on the remaining diffusion model. This approach is advantageous as it is currently not feasible to implement zkML on models as large as Stable Diffusion.

Based on the results obtained in Section 4.2, we can estimate the time and memory required for proof generation. The methodology is as follows: (1) we convert the four types of spatial transformers found within the U-Net component of Stable Diffusion into ONNX format; (2) we use the row counts from the EZKL-generated settings as an approximation of the final rows used when generating the proof; (3) we estimate the total rows needed for the full model using the average from the extrapolation of the four transformers (since the full model is too large for settings generation); (4) we predict the proof generation time and maximum memory usage by linear extrapolation from the previous results. The extrapolated data is available in Table 2.

Table 2: Stable Diffusion U-Net attention layers

| Model Layer | Input shape | No. of params | EZKL rows | Memory (TB)** | Time (hours)** |
|---|---|---|---|---|---|
| spatial_transformer | (64, 64, 320) | 2546240 (0.30%) | 10821242482 | 180 | 394 |
| spatial_transformer_2 | (32, 32, 640) | 9188480 (1.07%) | 5107529660 | 85 | 185 |
| spatial_transformer_4 | (16, 16, 1280) | 34760960 (4.04%) | 4429482857 | 74 | 161 |
| spatial_transformer_6 | (8, 8, 1280) | 34760960 (4.04%) | 1147973487 | 19 | 42 |
| Full model | - | 859520964 (100%) | 1067139337445*** | 17775 | 38842 |

It is estimated that the complete U-Net component of the Stable Diffusion model would require approximately 17,775TB of memory and more than four years to prove. However, individual transformers would only require 19-180TB of memory and 2-17 days to complete. With the availability of high-memory instances with up to 24TB of memory on cloud computing platforms like AWS*, opp/ai is bringing the possibility of model privacy on fine-tuned models based on large-scale open-source models much closer to reality in the near future.

---
* The PoC can be accessed at https://github.com/socathie/oppai-benchmark/tree/main/sd.
** extrapolated from Table 1
*** averaged and extrapolated from ratio between number of parameters and number of EZKL rows

## 5 DISCUSSIONS

### 5.1 Framework agnostic

While the benchmarking was only conducted on two zkML frameworks, the underlying principle of opp/ai applies to any framework. Any implementation of zkML can be combined with any implementation of opML to achieve similar reductions in computational costs. One recommended implementation is to integrate zkML closely with the current implementation of opML, which is developed under the *mlgo* [34] framework in the Go programming language. Compiling models written in *mlgo* with *zkGo* [29] into *WebAssembly* files that can be verified with *zkWasm* [30] could be a very convenient way to seamlessly prove any ML models, with the additional advantage of using *SoftFloat* [35] instead of integer/modular arithmetic, avoiding the need for model quantization.

### 5.2 Adapting to different types of ML models

The adaptability of our approach to various model architectures is a critical aspect of its utility. Single-path neural networks [20], due to their linear feedforward structure, are inherently more compatible with the opp/ai framework compared to models with multiple paths or branches. This distinction arises from the complexity associated with proving the correctness of computations in branched models, which may introduce additional challenges in generating and verifying zero-knowledge proofs, as well as the complexity of connecting its opML counterpart. However, this does not preclude the application of our methodology to complex models; rather, it underscores the need for tailored strategies to accommodate the unique characteristics of different model architectures.

### 5.3 Use cases

In Section 4.3, we have presented an example of privatizing the fine-tuned weights of an image generation model. These models can generate images based on textual descriptions, and the fine-tuned weights may contain proprietary enhancements that improve the model's performance or enable unique features. By utilizing the opp/ai framework, the model provider can compute the submodel proofs running in zkML, which allows the fine-tuned weights to remain private. The provider can then submit these proofs to on-chain verifiers, making the model available for use without revealing its proprietary aspects.

Another use case for the opp/ai framework is individual voice tuning in text-to-voice models. Text-to-voice service providers may offer personalized voice models that are tailored to the individual's voice characteristics. These personalized models are sensitive and contain valuable data. The opp/ai framework can ensure that the personalized voice model's parameters remain confidential while still offering the service to end-users verifiably.

In the financial sector, trading algorithms are developed to predict market movements and execute trades automatically. These algorithms are highly valuable and contain sensitive strategies that firms wish to protect. A financial institution could use the opp/ai framework to conceal the weights of a model that has been specifically tuned to its trading strategy. This allows the institution to use blockchain for secure, transparent, and verifiable trading operations without exposing the proprietary aspects of their models.

---

\* See https://aws.amazon.com/ec2/instance-types/high-memory.

In the gaming industry, AI models are used to create challenging and engaging non-player characters (NPCs). Game developers may fine-tune these models to create unique behaviors or strategies that are specific to their game. By using the opp/ai framework, developers can hide the fine-tuned weights that contribute to the NPCs' competitive edge, preventing other developers from copying these features while still providing an immersive gaming experience.

## 6 CONCLUSIONS

In conclusion, the opp/ai framework represents a significant advancement in the realm of blockchain-based machine learning, addressing the dual challenges of privacy and computational efficiency. By combining the privacy-preserving capabilities of zkML with the computational efficiency of opML, opp/ai offers a balanced solution that mitigates the cryptographic overhead while maintaining security through economic deterrents. The framework's adaptability to different models and its potential for various applications underscore its versatility. Despite the inherent trade-offs, opp/ai provides a promising avenue for secure, decentralized, and transparent AI services on the blockchain. The paper's analysis and benchmarking results affirm the framework's theoretical and practical efficacy, paving the way for its adoption and further development in the blockchain ecosystem.

## ACKNOWLEDGMENTS

We extend our deepest gratitude to Prof. Yupeng Zhang from the University of Illinois at Urbana-Champaign for his invaluable review and insightful advice on this paper, with a particular emphasis on the security analysis section. His expertise and thoughtful feedback have significantly enhanced the quality and depth of our work. We are truly appreciative of his contributions to our research.

## REFERENCES

[1] Suma Bhat, Canhui Chen, Zerui Cheng, Zhixuan Fang, Ashwin Hebbar, Sreeram Kannan, Ranvir Rana, Peiyao Sheng, Himanshu Tyagi, Pramod Viswanath, and Xuechao Wang. 2023. SAKSHI: Decentralized AI Platforms. *arXiv [cs.CR]* (July 2023). Retrieved from https://arxiv.org/abs/2307.16562

[2] Cao, Shen, Xie, Parkhi, and Zisserman. 2018. VGGFace2: A Dataset for Recognising Faces across Pose and Age. In *2018 13th IEEE International Conference on Automatic Face & Gesture Recognition (FG 2018)*, 67–74.

[3] K. D. Conway, Cathie So, Xiaohang Yu, and Kartin Wong. 2024. opML: Optimistic Machine Learning on Blockchain. *arXiv [cs.CR]*. Retrieved from http://arxiv.org/abs/2401.17555

[4] Creators Tensorflow Developers. *TensorFlow*. DOI:https://doi.org/10.5281/zenodo.10126399

[5] Muhammad ElSheikh and Amr M. Youssef. 2022. Dispute-free Scalable Open Vote Network using zk-SNARKs. *arXiv [cs.CR]*. Retrieved from http://arxiv.org/abs/2203.03363

[6] Jens Groth. 2016. On the Size of Pairing-based Non-interactive Arguments. *Cryptology ePrint Archive* (2016). Retrieved February 18, 2024 from https://eprint.iacr.org/2016/260.pdf

[7] Edward J. Hu, Yelong Shen, Phillip Wallis, Zeyuan Allen-Zhu, Yuanzhi Li, Shean Wang, Lu Wang, and Weizhu Chen. 2021. LoRA: Low-Rank Adaptation of Large Language Models. *arXiv [cs.CL]*. Retrieved from http://arxiv.org/abs/2106.09685

[8] Harry A. Kalodner, Steven Goldfeder, Xiaoqi Chen, S. Weinberg, and E. Felten. 2018. Arbitrum: Scalable, private smart contracts. *USENIX Security Symposium* (2018). Retrieved February 16, 2024 from https://www.semanticscholar.org/paper/e5559ea78af0685df47b534c4d96c5ae09474501

[9] Jong Hwan Ko, Taesik Na, Mohammad Faisal Amir, and Saibal Mukhopadhyay. 2018. Edge-Host Partitioning of Deep Neural Networks with Feature Space Encoding for Resource-Constrained Internet-of-Things Platforms. In *2018 15th IEEE International Conference on Advanced Video and Signal Based Surveillance (AVSS)*, IEEE, 1–6.

[10] Satish Kumar, Weng Marc Lim, Uthayasankar Sivarajah, and Jaspreet Kaur. 2023. Artificial Intelligence and Blockchain Integration in Business: Trends from a Bibliometric-Content Analysis. *Inf. Syst. Front.* 25, 2 (2023), 871–896.

[11] Y. Lecun, L. Bottou, Y. Bengio, and P. Haffner. 1998. Gradient-based learning applied to document recognition. *Proc. IEEE* 86, 11 (November 1998), 2278–2324.

[12] Dongho Lee, Jaeseo Lee, Hyunsung Jun, Hongdeok Kim, and Seehwan Yoo. 2021. Triad of Split Learning: Privacy, Accuracy, and Performance. In *2021 International Conference on Information and Communication Technology Convergence (ICTC)*, IEEE, 1185–1189.

[13] Omkar M. Parkhi, A. Vedaldi, and Andrew Zisserman. 2015. Deep Face Recognition. *Br Mach Vis Conf* (2015), 41.1-41.12.

[14] P. Peduzzi, J. Concato, E. Kemper, T. R. Holford, and A. R. Feinstein. 1996. A simulation study of the number of events per variable in logistic regression analysis. *J. Clin. Epidemiol.* 49, 12 (December 1996), 1373–1379.

[15] Ethereum Improvement Proposals. 2023. ERC-7007: Zero-Knowledge AI-Generated Content Token. *Ethereum Improvement Proposals*. Retrieved February 17, 2024 from https://eips.ethereum.org/EIPS/eip-7007

[16] Robin Rombach, Andreas Blattmann, Dominik Lorenz, and Patrick Esser Bjorn Ommer. High-Resolution Image Synthesis with Latent Diffusion Models. Retrieved from https://github.com/CompVis/latent-diffusion

[17] Imam Santoso and Yuli Christyono. 2023. Zk-SNARKs as A cryptographic solution for data privacy and security in the digital era. *Computational* 12, 2 (August 2023), 53–58.

[18] Scroll. 2023. The Cost of Intelligence - Ryan Cao (Modulus Labs). (February 2023). Retrieved February 9, 2024 from https://www.youtube.com/watch?v=nsMM2iE_oUA

[19] Cathie So. *keras2circom: python tool to transpile a tf.keras model into a circom circuit*. Github. Retrieved February 18, 2024 from https://github.com/socathie/keras2circom

[20] Surat Teerapittayanon, Bradley McDanel, and H. T. Kung. 2017. Distributed Deep Neural Networks Over the Cloud, the Edge and End Devices. In *2017 IEEE 37th International Conference on Distributed Computing Systems (ICDCS)*, IEEE, 328–339.

[21] Florian Tramèr, Fan Zhang, Ari Juels, Michael K. Reiter, and Thomas Ristenpart. 2016. Stealing Machine Learning Models via Prediction APIs. *arXiv [cs.CR]*. Retrieved from http://arxiv.org/abs/1609.02943

[22] Daniel Davis Wood. 2014. ETHEREUM: A SECURE DECENTRALISED GENERALISED TRANSACTION LEDGER. (2014). Retrieved from https://ethereum.github.io/yellowpaper/paper.pdf

[23] Zihan Zheng, Peichen Xie, Xian Zhang, Shuo Chen, Yang Chen, Xiaobing Guo, Guangzhong Sun, Guangyu Sun, and Lidong Zhou. 2021. Agatha: Smart Contract for DNN Computation. *arXiv [cs.CR]*. Retrieved from http://arxiv.org/abs/2105.04919

[24] 2023. An introduction to zero-knowledge machine learning (ZKML). Retrieved February 16, 2024 from https://worldcoin.org/blog/engineering/intro-to-zkml

[25] 2023. Privacy at Worldcoin: A Technical Deep Dive - Part I. Retrieved February 18, 2024 from https://worldcoin.org/blog/developers/privacy-deep-dive

[26] *vggface: @rcmalli's keras-vggface library updated to Tensorflow 2*. Github. Retrieved February 17, 2024 from https://github.com/YaleDHLab/vggface

[27] *circom: zkSnark circuit compiler*. Github. Retrieved February 17, 2024 from https://github.com/iden3/circom

[28] *halo2*. Github. Retrieved February 17, 2024 from https://github.com/privacy-scaling-explorations/halo2

[29] *go at zkGo*. Github. Retrieved February 17, 2024 from https://github.com/ethstorage/go

[30] *zkWasm*. Github. Retrieved February 17, 2024 from https://github.com/DelphinusLab/zkWasm

[31] *ezkl: ezkl is an engine for doing inference for deep learning models and other computational graphs in a zk-snark (ZKML). Use it from Python, Javascript, or the command line*. Github. Retrieved February 18, 2024 from https://github.com/zkonduit/ezkl

[32] Simple MNIST convnet. Retrieved February 18, 2024 from https://keras.io/examples/vision/mnist_convnet/

[33] *snarkjs: zkSNARK implementation in JavaScript & WASM*. Github. Retrieved February 18, 2024 from https://github.com/iden3/snarkjs

[34] *mlgo*. Github. Retrieved February 18, 2024 from https://github.com/OPML-Labs/mlgo

[35] *berkeley-softfloat-3: SoftFloat release 3*. Github. Retrieved February 18, 2024 from https://github.com/ucb-bar/berkeley-softfloat-3

# A APPENDICES

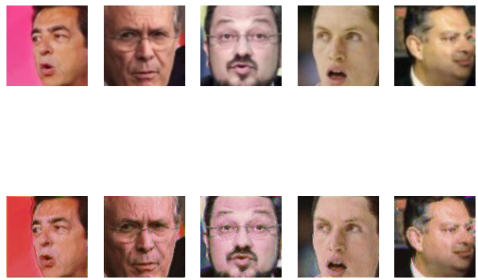

Figure A.1: The results of a reconstruction attack trained on the inference outputs from the first few layers of the VGGFace model. The top row displays the original images, while the bottom row displays the reconstructed images.*

---

\* The complete simulation script can be accessed at https://github.com/socathie/oppai-benchmark/tree/main/vggface_inverse.